\documentclass{article}
\newcommand{\be}{\begin{equation}}
\newcommand{\ee}{\end{equation}}
\newcommand{\bold}{\textbf}

\begin{document}

%\title{\LARGE{\bold{Unified Description of Plausible Cause
%and Effect \\ of the Big Bang}}\thanks{ To be published by the
%American Institute of Physics in the Proceedings of the
%University of California campus in Berkeley, California on July
%30-August 3, 2001.}}
%\author{\Large{D.C. Choudhury} \\ \normalsize{Department of Physics, Polytechnic University,
%Brooklyn, New York 11201 USA} \\
%\normalsize{e-mail:  dchoudhu@duke.poly.edu}}
%\date{\normalsize{14 November 2001}}
%\maketitle

\begin{center}\LARGE \textbf{Resolution of Cosmological
Singularity and a Plausible Mechanism of the Big Bang}\footnote{Talk given at INPC 2001. To
be published by the American Institute of Physics in the
Proceedings of the International Nuclear Physics Conference that
took place at the University of California campus in Berkeley,
California on July
30-August 3, 2001.} \\  \normalsize \vspace{2em} \Large{D.C. Choudhury} \\
 \normalsize{Department of Physics, Polytechnic University,
Brooklyn, New York 11201 USA
\\ e-mail: dchoudhu@duke.poly.edu}
\end{center}

\begin{quote}
\textbf{Abstract.}  The initial cosmological singularity in the
framework of the general theory of relativity is resolved by
introducing the effect of the uncertainty principle of quantum
theory without violating conventional laws of physics. A plausible
account of the mechanism of the big bang, analogous to that of a
nuclear explosion, is given and the currently accepted Planck
temperature of $\sim 10^{32}$~K at the beginning of the big bang
is predicted.

\vspace{1ex} Subj-class:  cosmology:  theory-pre-big bang;
mechanism of the big bang.
\end{quote}

\vspace{2em}

\normalsize \emph{I. Introduction.}  The standard hot Big Bang
model within the framework of the general theory of relativity has
been spectacularly successful in describing the evolution of the
universe from ten thousandth of a second after the big bang to the
present time. Nevertheless, the occurrence of the initial
singularity, a state of infinite density and temperature,
preceding the big bang as shown by the theorems [1,2] of Penrose
and of Hawking and Penrose, raises some of the deepest questions
concerning the foundation of this model of cosmology. To resolve
this problem, we propose that during the process of the
gravitational collapse of the universe, in the domain of the
infinitesimally small region of space-time around the initial
singularity, the general theory of relativity by itself is
incomplete and the introduction of the effect of quantum theory
becomes absolutely necessary. Therefore it is assumed that at the
end of the gravitational collapse, without violating the
uncertainty principle, the total mass of the universe including
all forms of energy condensed into the primordial black hole (PBH)
in a state of $\sim$ absolute zero temperature. The reasoning for
this assumption is consistent with Penrose's argument based on
thermodynamic considerations that the universe was initially very
regular (i.e., in a state of zero entropy); Chandrasekhar [3]
likewise agreed with Penrose's view, contrary to those who hold
that the chaos at the initial singularity was maximal.

\emph{II. Quantization of energy.}  The energy of the collapsed
state of the universe (PBH), in light of the above argument, is
quantized in such a manner that each quantum of mass (frozen
energy) is characterized by the fundamental constants of the
general theory of relativity and of quantum theory (namely $G,
\hbar, c$).  For this purpose the principle of dimensional
analysis and the reciprocity relation, analogous to that of the
unitarity condition of quantum theory, are incorporated into our
method of quantization. Consequently, the physical properties of
the quantized energy (cosmic particle) remain independent of the
units in which they are measured and also the magnitude of its
mass has the lowest value. Since the dimensions of $G, \hbar, c$,
and mass ($M$) are known, the only simplest dimensionless
combinations which can be formed of them are $GM^2/\hbar c$ and
$\hbar c/ GM^2$.  Therefore they satisfy the following condition:

\be \frac{\hbar c}{GM^2} = \frac{GM^2}{\hbar c}. \ee

The only physical solution out of the four solutions, $\pm (\hbar
c/G)^{1/2}$ and $\pm i (\hbar c/G)^{1/2}$ for $M$, we have chosen
the real positive  value for the mass $M$:

\be M=\left( \frac {\hbar c}{G} \right)^{1/2} \cong 2.2 \times
10^{-5}~\mathrm{gm}. \ee

This is the lowest mass of the quantized frozen energy which we
call the cosmic particle. The magnitude of this mass is identical
to that of the Planck's mass which he derived from his theory of
the black body radiation (Planck 1899) [4].  Although the values
of their masses are the same, their physical properties are
entirely different. One is an elementary quantum of frozen energy
which has only rest mass while the other is the largest quantum of
radiation energy which has zero rest mass. The intrinsic spin of
this cosmic particle is considered to be zero because: (i) The
method of quantization of frozen energy is independent of any spin
and depends only on the constants $G, \hbar$, and $c$; ;   (ii)
The method of quantization does not involve any dynamical operator
which can generate intrinsic spin;  (iii) If this particle had an
intrinsic spin, it would not have the lowest cosmic unit of mass,
rather its lowest mass would be increased by an amount equivalent
to the rotational energy of its intrinsic spin; and finally (iv)
Its shape must be perfectly spherical because it is in  a state of
absolute zero temperature and of lowest energy state. Therefore,
we conclude that most likely its intrinsic spin is zero and it is
a boson.

\emph{III. Properties of quantized particles.}  The minimum size
and minimum lifetime of the quantized cosmic particles, each of
mass, $M = (\hbar c/G)^{1/2}$, are determined within the framework
of the uncertainty principle (Heisenberg 1927) [5]. Profound
implications of the uncertainty principle in physical terms are
presented in complementarity principle (Bohr 1928) [6].  In a
four-dimensional space-time, coordinates of a particle and its
corresponding components of energy-momentum, the uncertainty
relations are:  \[ \Delta x \cdot \Delta p_x \geq \hbar ; \ \
\Delta y \cdot \Delta p_y \geq \hbar ; \ \ \Delta z \cdot \Delta
p_z \geq \hbar \ \ \mathrm{and} \ \Delta t \cdot \Delta E \geq
\hbar . \] \[ \mathrm{Let} (\Delta x)_{min} = (\Delta y)_{min} =
(\Delta z)_{min} \equiv L_{min} , \]
\[(\Delta p_x)_{max} = (\Delta p_y)_{max} = (\Delta p_z)_{max}
\equiv P_{max} = Mc,\]
\[(\Delta t)_{min} \equiv \tau_m \ \mathrm{and} \ (\Delta E)_{max}
\equiv E = Mc^2, \] therefore: \be L_{min} \cong \frac {\hbar}{Mc}
= \left( \frac {\hbar G}{c^3} \right) ^{1/2} \approx \ 1.6 \times
10^{-33}~\mathrm{cm~and} \ee \be \tau_m \cong
\left(\frac{\hbar}{Mc^2}\right) = \left( \frac{\hbar
G}{c^5}\right) ^{1/2} \approx \ 5.3\times 10^{-44}~\mathrm{sec}.
\ee

$L_{\mathrm{min}}(L)$ and $\tau_m$ are interpreted as the minimum
size of localization and minimum lifetime (time scale) of the
quantized cosmic particle. It is significant that the values of
Planck's mass, length, and time (Planck 1899) [4] are identical to
those of the mass, size, and minimum lifetime of the cosmic
particle obtained in the present investigation. However, they are
entirely different in their properties and these different
physical properties of the cosmic particle play the most important
role in the big bang theory. Thus our results provide new
significance into Planck's system of units and show that these
units are fundamental.

\emph{IV. Structure of the modified singularity.} The constituents
of the modified initial cosmological singularity (PBH) are cosmic
particles, each of mass, $M = (\hbar c / G)^{1/2}$, and minimum
size of localization, $L_{\mathrm{min}} = (G \hbar /c^3)^{1/2}$.
These particles are bosons and obey the Bose-Einstein statistics
(Bose 24; Einstein 1924,1925a) [7,8]. Consequently, without
violating the uncertainty principle of quantum theory and putting
all the Bose-Einstein particles in the lowest energy state as $T
\rightarrow 0$, the maximum density of matter,
$\rho_{\mathrm{max}}$, in the primordial black hole (PBH) is given
as a function of $G, \hbar$, and $c$:  \be \rho_{max} = \frac
{3}{4\pi} \left(\frac{c^5}{\hbar G^2}\right) \cong 1.2\times
10^{93} ~\mathrm{gm/cm^3}. \ee

The minimum volume and the minimum size ($r_{\mathrm{min}}$) of
the structure of the modified initial singularity are determined
taking the total mass of the universe to be about $5.68 \times
10^{56}$~gms.  This estimation of mass is based on a typical
cosmological model without cosmological constant, compatible with
astronomical observations and with Einstein's conception of
cosmology (Misner, Thorne, \& Wheeler 1973) [9].  We have already
estimated the mass density to be about $10^{93}$~gm/cm$^3$ which
together with the total mass of the universe leads to the minimum
volume $4.76 \times 10^{-37}$~cm$^3$ and the minimum size of the
radius ($r_{\mathrm{min}}$) $4.8 \times 10^{-13}$~cm, about one
hundred thousandth of the size of an atom.  These results are
significant because they provide a physical interpretation and a
reasonable resolution of the initial cosmological singularity
predicated by the general theory of relativity. It is relevant to
point out here that these numbers (related to minimum volume and
minimum radius) may be considered illustrative because they depend
on the mass of the gravitationally collapsed state. Nevertheless
the present procedure, within the framework of the general theory
of relativity and of the uncertainty principle of quantum theory,
remains valid to resolve the problem of the initial cosmological
singularity. The preservation of the laws of thermodynamics are
discussed below.

\emph{V. Thermodynamic properties of the PBH.}  From the available
information about the temperatures of white dwarfs, neutron stars,
and black holes (Hawking 1988) [10] , and as well as from
Penrose's argument (Chandrasekhar 1990) [3] we infer,  as  stated
above, that the temperature of the PBH goes to zero in its final
collapsed state. To calculate the entropy of this system we
utilize Boltzmann's relation between the entropy $S$ and
thermodynamic probability $W$: \be S = k_\beta \ln W + S_\circ \ee
where $k_{\beta}$ and $S_\circ$ are Boltzmann's and integration
constants respectively. The first term, $k_\beta \ln W$ is
interpreted as the internal entropy of the matter present inside
the PBH and $S_\circ$ as the external entropy but within the
boundary of the surface area of the event horizon which is
inaccessible to an exterior observer. Since the quantized matter
inside the PBH consists of bosons, its properties can be described
by Bose-Einstein statistics. Therefore the probability $W$ can be
written as (Einstein 1924,1925a) [8]: \be W = \prod_{s}
\frac{\left( N^s +Z^s -1 \right)!}{N^s! \left(Z^s - 1\right)!}.\ee
where $N^s$ are the number of particles and $Z^s$ are the number
of subcells in the $s$-th cell. We have pointed out above, all $N$
bosons go into the lowest energy state as $T \rightarrow 0$, in
this limit $N^s=0$ for all other cells. Therefore $W \rightarrow
1$, $k_\beta \ln W \rightarrow 0$, and $S \rightarrow S_\circ$.
Consequently the internal entropy of PBH $\rightarrow 0$ as $T
\rightarrow 0$.   This result is consistent with the prediction of
Einstein in 1925 that B-E gas satisfies the third law of
thermodynamics (Einstein 1925b; Pais 1982) [11,12].

The entropy $S_\circ$, inaccessible to an exterior observer, is
calculated within the framework of the theory of black hole
entropy based on the second law of thermodynamics and information
theory developed by Bekenstein (1973) [13]. Therefore:  \be S =
S_\circ = \frac{1}{2} \left(\frac{\ln 2}{4\pi} \right) k_\beta c^3
\hbar^{-1} G^{-1} A \ee where $A$ is the surface area of the event
horizon of the PBH corresponding to the Schwarzschild's radius
$R_{\mathrm{Sch}} = 2GM/c^2$ and $A = 16 \pi M^2 G^2/c^4$; here
$M$ represents the total mass of PBH. The estimated value of the
entropy of PBH is found to be $\approx 1.32 \times
10^{107}$~erg$\cdot$K$^{-1}$ which is extremely large as expected.

Finally the internal pressure in the final phase of the
gravitational collapse (PBH) is evaluated within the framework of
correspondence principle (Bohr 1923) [14] and Newton's
gravitational theory, assuming that the pressure inside the PBH is
generated by an immense, attractive gravitational force alone. At
a distance $r$ from the center of PBH, the pressure $P(r)$ is
given by:  \be P(r) = \frac{2\pi}{3}G\rho_{max}^2 \left( r_{min}^2
- r^2 \right); \ \ r \leq r_{min}, \ee where $\rho_{\mathrm{max}}$
and $r_{\mathrm{min}}$ are the maximum matter density and minimum
radius of the (PBH) respectively. The maximum pressure when $r
\rightarrow 0$, at the center, $P(r \rightarrow 0)$ is calculated
using the earlier estimated values of $\rho_{\mathrm{max}} \approx
1.2 \times 10^{93}$~gm/cm$^3$ and $r_{\mathrm{min}} \approx 4.8
\times 10^{-13}$~cm.  The result is $P(r \rightarrow 0) \approx 5
\times 10^{154}$~dynes/cm$^2$, indeed extremely large.

\emph{VI. Plausible mechanism of the big bang.}  Based on our
knowledge of nuclear fission and nuclear explosives, we give a
simplified plausible account of the mechanism of the big bang.
Just as the salient features of the fission of a nucleus, say
uranium by slow neutrons, have been fully interpreted in terms of
the liquid drop model, originally by Bohr \& Wheeler (1939) [15];
subsequently, this information has been successfully utilized in
the design of various types of explosives -- from nuclear fission
to nuclear hydrogen bombs. For example, the basic design of a
nuclear implosion bomb (Krane 1987) [16] consists of a solid
spherical subcritical mass of the fissionable material (e.g.,
$^{235}$U or $^{239}$Pu) surrounded by a spherical shell of
conventional chemical explosives.  An initiator at the center
provides neutrons to start a chain reaction. When the conventional
explosives are detonated in exact synchronization, a spherical
shock wave compresses the fissionable material into a supercritical
state, resulting in an explosion. Now we illustrate that many of
the basic characteristics of a nuclear fission bomb are naturally
present in PBH.

In the present instance: (1) the PBH which consists of unstable
quantized cosmic particles (bosons) is analogous to the solid
spherical subcritical mass of the fissionable material;  (2) the
vacuum energy fluctuations caused by the uncertainty principle of
quantum theory surrounding the PBH (within the boundary of its
event horizon) can produce results similar to those of the
conventional explosives surrounding the fissionable material of
the bomb; (3) the unstable cosmic particles, each of Planck's
mass, are analogous to the nuclei of the fissionable material; and
(4) the constraint of uncertainty principle and the maximum
gravitational  pressure at the innermost  central  region of  PBH
may trigger the explosion (big bang) analogous to  that  of  the
initiator at the center of the bomb which provides neutrons to
begin the chain reaction resulting in explosion. In an explosion
of a nuclear bomb, only a very small fraction of mass of each
nucleus of fissionable material is transformed into kinetic energy
while in the explosion of the PBH, total mass ($M \cong 2.2 \times
10^{-5}$~gms) of each unstable cosmic particle is converted into
kinetic energy resulting in temperature of $10^{32}$~K, the
currently accepted temperature at the beginning of the big bang.
The total mass of the PBH consisting of nearly $10^{61}$ cosmic
particles, each of mean lifetime (decay rate) of the order of
$\tau_m \approx 5.3 \times 10^{-44}$~sec, is transformed into
kinetic energy $\approx 10^{77}$~erg, in less than
$10^{-40}$~second resulting in the biggest explosion, the big
bang.

The extension and implications of the present investigation are
offered in Ref. [17] for the large-scale structure of the universe
within the framework of the Friedmann-Robertson-Walker cosmology
[18] also known as the standard cosmology.

\emph{VII. Conclusion.}  The initial singularity, a state of
infinite density and temperature preceding the big bang, raises
grave questions concerning the foundation of the big bang model of
the modern cosmology. This problem is resolved by introducing the
effects of the uncertainty principle in the domain of the
infinitesimally small region of space-time during the process of
the gravitational collapse of the universe. Consequently the end
of the collapse, instead of resulting in a state of singularity,
results in the primordial black hole (PBH), a state of finite
volume, finite density, finite temperature $\rightarrow 0$, and
the finite internal entropy of the matter present inside the PBH
$\rightarrow 0$. A plausible mechanism of the big bang, an
explosion of the PBH, analogous to that of a nuclear explosion, is
presented. The justification for the basic assumptions in the
present investigation is also given within the framework of the
conventional laws of physics.

\newpage
%\vspace{1em}
\flushleft \textbf{REFERENCES} \vspace{0.5em}

\begin{enumerate}
    \item[[1\hspace{-1ex}]] Penrose, R., \emph{Phys. Rev. Letters} \bold{14},
        57-59 (1965).
    \item[[2\hspace{-1ex}]] Hawking, S. W., and Penrose, R., \emph{Proc. Roy.
        Soc. London} \bold{A314}, 529-548 (1970).
    \item[[3\hspace{-1ex}]] Chandrasekhar, S., \emph{Selected Papers, Volume 5,
        Relativistic Astrophysics}, University of Chicago,
        Chicago, 1990, p. 577.
    \item[[4\hspace{-1ex}]] Planck, M., \emph{Sitzungsberichte, Deut. Akd. Wiss.
        Kl, Math-Phys. Tech.}, Berlin, 440-480 (1899).
    \item[[5\hspace{-1ex}]] Heisenberg, W., \emph{Zeits. f. Physik}, \bold{43},
        172-198 (1927).
    \item[[6\hspace{-1ex}]] Bohr, N., \emph{Nature}, \bold{121}, 580- 590 (1928).
    \item[[7\hspace{-1ex}]] Bose, S. N., \emph{Zeits, f. Physik}, \bold{26},
        178-181 (1924).
    \item[[8\hspace{-1ex}]] Einstein, A., \emph{Sitzb. Preuss. Akad. Wiss. Phys.-Math.
        Kl.}, Berlin, 261-267 (1924); ibid., 3-14 (1925a).
    \item[[9\hspace{-1ex}]] Misner, C. W., Thorne, K. S., Wheeler, J. W.,
        \emph{Gravitation}, W. H. Freeman \& Co., San Francisco,
        1973, p. 738.
    \item[[10\hspace{-1ex}]] Hawking, S. W., \emph{A Brief History of Time}, Bantam
        Books, New York, 1988,  pp. 105-108.
    \item[[11\hspace{-1ex}]] Einstein, A., \emph{Sitzb. Preuss. Akad. Wiss.
        Phys.-Math. Kl.} Berlin,  18-25  (1925b)
    \item[[12\hspace{-1ex}]] Pais, A., \emph{Subtle is the Lord: The Science and the Life of
        Albert Einstein}, Oxford University Press, New York, 1982, p. 431.
    \item[[13\hspace{-1ex}]] Bekenstein, J. D., \emph{Phys. Rev.} \bold{D7}, 2333-2346 (1973).
    \item[[14\hspace{-1ex}]] Bohr, N., \emph{Zeits. f. Physik} \bold{13}, 117-165
        (1923).
    \item[[15\hspace{-1ex}]] Bohr, N., and Wheeler, J. W., \emph{Phys. Rev.} \bold{56}, 426-450 (1939).
    \item[[16\hspace{-1ex}]] Krane, K. S., \emph{Introductory Nuclear Physics}, John Wiley
        \& Sons , Inc., New York, 1987, p. 521.
    \item[[17\hspace{-1ex}]] Choudhury, D. C., arXiv: astro-ph/0103456 (2001).
    \item[[18\hspace{-1ex}]] Kolb, E. W., and Turner, M. S., \emph{The Early Universe, Paper Back
        Edition}, Addison-Wesley, Massachusetts, 1994, pp. 47-86.
\end{enumerate}

\end{document}